\newcommand{\beq}{\begin{eqnarray}}
\newcommand{\eeq}{\end{eqnarray}}

\newcommand{\dX}{\frac{d{\bf X}}{dt}}
\newcommand{\du}{\frac{d{\bf u}}{dt}}
\newcommand{\bfu}{{\bf u}}
\newcommand{\bfx}{{\bf x}}
\newcommand{\bfU}{{\bf U}}
\newcommand{\bfX}{{\bf X}}
\newcommand{\bfF}{{\bf F}}
\newcommand{\bfW}{{\bf W}}
\newcommand{\tbfX}{\tilde{\bf X}}
\newcommand{\eps}{\epsilon}
\newcommand{\half}{{1\over 2}}
\newcommand{\e}{{\rm e}}

\renewcommand{\d}{\partial}
\renewcommand{\Re}{{\rm Re}}

\renewcommand{\theequation}{\thesection.\arabic{equation}}

\newcommand{\cl}{\centerline}

\newcommand{\btem}{\bibitem}

\newcommand{\PRL}{Phys.\ Rev. \ Lett.}

\documentstyle[12pt]{article}    
\begin{document}
\begin{flushright}
RYUTHP-96/1 \\
Feb., 1996 \\
Sep.(Rev.)\\ 
\end{flushright}

\begin{center}
{\large {\bf The Renormalization-Group  Method }}\\
{\large {\bf Applied to Asymptotic Analysis of Vector Fields}}
\end{center}

\vspace{1cm}

{\cl {Teiji Kunihiro}}

\bigskip

\begin{center}

Faculty of Science and Technology, Ryukoku University,\\ 
Seta, Ohtsu, 520-21, Japan\\ 
\end{center}

\begin{abstract}
The  renormalization group  method of Goldenfeld, Oono
 and their collaborators  is applied to asymptotic analysis of
vector fields. The method is formulated 
on the basis of the theory of envelopes, as was  done for scalar fields.
This formulation actually completes the discussion of the previous 
work for scalar equations. 
It is shown in a generic way that the method applied to 
 equations with a  bifurcation leads to the Landau-Stuart and 
 the (time-dependent) Ginzburg-Landau equations.
  It is confirmed that  this method 
  is actually   a powerful theory for the reduction of the dynamics 
as the reductive perturbation method is.  
Some examples  for ordinar diferential equations,
such as the forced Duffing, the Lotka-Volterra and the Lorenz 
equations, are worked out in this method: The time evolution of the
solution of the Lotka-Volterra equation is explicitly given, while the
 center manifolds of the Lorenz equation  are constructed in a simple 
way in the RG method.
\end{abstract}

\setcounter{equation}{0}
\section{Introduction}
\renewcommand{\theequation}{\thesection.\arabic{equation}}

It is well known that the renormalization group (RG) equations \cite{rg}
have a peculiar power to improve the global nature of functions obtained 
in the perturbation theory in quantum field theory (QFT)\cite{JZ}:
 The RG equations may be interpreted as representing  the fact that 
the physical quantities ${\cal O}(p,\alpha, \mu)$ 
should not depend on the renormalization point $\mu$
  having any arbitrary value,
\beq
\frac{\partial {\cal O}(p, \alpha;\mu)}{\d \mu}=0.
\eeq 
Such a floating renormalization
 point was first introduced by Gell-Mann and Low in the celebrated 
paper\cite{rg}.

  It is  Goldenfeld, Oono and their collaborators ( to be abbreviated
 to GO)
 \cite{goldenfeld1,goldenfeld2} who first showed that the RG 
equation can be used  for purely mathematical problems as 
 to improving the global nature of the solutions of differential 
 equations  obtained in the perturbation  theory.
 One might say, however, that their presentation of the method is
 rather heuristic, heavily relied on the RG prescription 
 in QFT and statistical physics; it seems that they were not so 
 eager to give a mathematical reasoning to the method so that  their 
 method may be  understandable even  for those who are not familiar with the
 RG.\footnote{In Appendix A, we give a brief account of the 
 Goldenfeld et al's prescription.}
  In fact, the reason why the RG equations even in QFT ``improve'' 
 naive perturbation had not been elucidated. One may say that when 
 GO successfully applied the
 RG method to purely mathematical problems such as solving differential
 equations, it had shaped a clear problem to reveal the mathematical 
 reason of the powefullness of the RG method,  at least, a la 
Stuckelberg-Peterman and Gell-Mann-Low.

Quite recently, the present author has formulated the method and 
 given the reasoning of GO's method on the basis of the classical theory of 
envelopes\cite{kunihiro,kunihiro2}:
 It was demonstrated that owing to the very RG equation,
the functions consturcted from the solutions in the perturbation theory 
 certainly satisfies the differential equation in question uniformly up 
 to the order with which local solutions around $t=t_0$ is constructed.
It was  also indicated in a generic way that 
the RG equation may be regarded as  the envelope equation. 
In fact, if  a family of  curves $\{{\rm C}_{\mu}\}_{\mu}$ 
in the $x$-$y$ plane  is  represented by $y=f(x; \mu)$,
  the function $g(x)$ representing the envelope E is given  
by eliminating the parameter $\mu$ from the  equation
\beq
\frac{\partial f(x; \mu)}{\partial \mu}=0.
\eeq
One can readily recognize the similarity of the envelope equation 
Eq.(1.2) with the  RG equation Eq.(1.1).
In Ref.'s\cite{kunihiro,kunihiro2}, 
 a simplified prescription  of the RG method is also presented.
 For instance, the perturbative expansion is made with 
respect to a small parameter and independent functions\footnote{Such an
 asmptotic series is called {\em generalizes asymptotic series}.
 The author is gratefull to T. Hatusda for telling him this fact and 
 making him recognize its
 significance.}, and
 the procedure of the ''renormalization"  has been 
shown unnecessary. 

However, the work given in \cite{kunihiro,kunihiro2} may be said to be
  incomplete
 in the following sense:
 To give the proof mentioned above, the scalar equation in question 
 was converted to a
 system of {\em first order} equations, which describe a vetor field.
But the theory of envelopes for vetor fields, i.e.,envelopes of 
trajectories, has not been presented in \cite{kunihiro,kunihiro2}. 
 The theory should have
 been formulated for vector equations to make the discussion  
 self-contained and complete.
 
One of the purposes of the present paper is therefore 
to reformulate geometrically the RG
 method for  vector equations, i.e., systems of ODE's and PDE's and
 to complete the discussion given in \cite{kunihiro,kunihiro2}.

Another drawback of the previous work is that
 a reasoning given to a procedure to setting $t_0=t$
 in the RG method\footnote{See Appendix A.} of Goldenfeld et al
 was not fully persuasive.\footnote{The author is  grateful to Y. Oono
 for his criticism on this point.} In this paper, we present  a more
 convincing reasoning for the procedure.

Once the RG method is formulated for vector
 fields, 
the applicability of the 
RG method developed by Goldenfeld, Oono and their collaborators
 is found to be wider than one might have imagined: 
The RG method is applicable also to,
 say,$n$-dimensional vector equations
 that are not simply converted to a scalar equation of the  $n$-th 
order; needless to say, it is not necessarily possible to convert
 a system of ordinary differential equations (or dynamical system)  
to a scalar equation of a high order with a simple structure,
 although the converse is always possible
 trivially.
 For partial differential equations,
it is not always possible to convert a system to a scalar equation of a
 high order\cite{curant}.
 Moreover, interesting equations in science including physics and 
applied mathematics
 are often given as a system. Therefore, it is of interest and 
importance to show that the RG method can be extended and applied to 
vector equations. To demonstrate the powefulness of the method,
 we shall work out  some specific examples of vector equations.

We shall emphasize that the RG method provides 
 a general method for the reduction of the dynamics as the
 reductive perturbation method (abbreviated to RP method)\cite{kuramoto} 
 does. 
 It should be mentioned that Chen, Goldenfeld and Oono\cite{goldenfeld2}
 already indicated that it is a rule that  the 
RG equation  gives equations for slow motions which the RP method may 
also  describe.  
In this paper, we shall confirm their observation 
in a most general setting  for vector equations.
Furthermore, one can show \cite{kunihiro3} that the natural extension of 
the RG method also applies to {\em difference} equations or maps, and 
an extended envelope equation leads to a reduction of the dynamics
 even for discrete maps.  
Thus one sees that the RG method is truly 
 a most promising candidate for a general theory of 
the reduction of  the dynamics, although actual computation is often
 tediuous in such a general and mechanical method.
 
This paper is organized as follows:
In the next section, we desribe the  theory of envelopes for curves (or
 trajectories) in 
 parameter representation. 
In section 3, the way how to construct envelope surfaces is given when
 a family of surfaces in three-dimensional space are parametrized 
 with two parameters.
  In section 4, we give the
 basic mathematical theorem for the RG method applied to
  vector fields.  
 This section is  partially a recapitulation of
 a part of Ref.\cite{kunihiro}, although some 
clarifications are made here.
In section 5, some examples are  examined in this method,
 such as 
 the forced Duffing\cite{holmes}, the Lotka-Volterra\cite{lotka}
 and the Lorenz\cite{lorenz,holmes} equations. 
 The Duffing equation is also an example of non-autonomous one,
  containing an external force.
 In section 6, we treat  generic equations with a bifurcation;
the Landau-Stuart\cite{stuart} and 
the Ginzburg-Landau equation will be derived in the RG method.
The final section is devoted to a brief summary and concluding remarks.
In Appendix A, a critical review of the Goldenfeld et al's  method is 
given.
In Appendix B, the Duffing equation is solved as a scalar equation
in the  RG method.
	
\section{Envelopes of trajectories}
\setcounter{equation}{0}
\renewcommand{\theequation}{\thesection.\arabic{equation}}

 To give a geometrical meaning to the RG equation for systems, one
 needs to formulate a theory of envelopes of curves which are 
 given in a parameter representation: For example,
 if the equation is for 
$\bfu (t)=\ ^t(x(t), y(t))$, the solution forms a trajectory or curve in the
 $x$-$y$ plane with $t$ being a parameter.
In this section, we give a brief account of the classical theory 
of envelopes for curves in the $n$-dimensional space, 
given in a parameter representation.

Let a family of curves $\{{\rm C}_{\alpha}\}_{\alpha}$ in  an 
$n$-dimensional space be given by 
\beq
\bfX (t; \alpha)=\ ^t(X_1(t; \alpha), X_2(t; \alpha), ... ,
X_n(t;\alpha)),
\eeq
where the point $(X_1, X_2,... X_n)$ moves in the $n$-dimensional space
 when $t$ is varied. Curves in the family is parametrized by $\alpha$.
We suppose that the family of curves $\{{\rm C}_{\alpha}\}_{\alpha}$
 has the envelope E:
\beq
\bfX _E(t)=\ ^t(X_{E1}(t; \alpha), X_{E2}(t; \alpha), ... ,X_{En}
(t;\alpha)).
\eeq

The functions $\bfX _E(t)$ may be obtained from $\bfX(t;\alpha)$ as 
follows. If the contact point of C$_{\alpha}$ and E is given by 
$t=t_{\alpha}$, we have 
\beq
\bfX(t_\alpha; \alpha)=\bfX_E(t_{\alpha}).
\eeq
For each  point in E, 
there exists a parameter $\alpha=\alpha(t)$: Thus the envelope 
function is  given by 
\beq
\bfX_E(t_{\alpha})=\bfX(t_\alpha; \alpha(t_{\alpha})).
\eeq

Then the problem is to get the function $\alpha(t)$, which is achieved
 as follows.
The condition that E and C$_{\alpha}$ has the common tangent line at
$\bfX(t_\alpha; \alpha)=\bfX_E(t_{\alpha})$ reads
\beq
\frac{d\bfX}{dt}\biggl\vert_{t=t_{\alpha}}=
\frac{d\bfX_E}{dt}\biggl\vert_{t=t_{\alpha}}.
\eeq
On the other hand, differentiating Eq.(2.4), one has
\beq
\frac{d\bfX_E}{dt}\biggl\vert_{t=t_{\alpha}}=\frac{\d \bfX}{\d t}
\biggl\vert_{t=t_{\alpha}}+ \frac{\d \bfX}{\d \alpha}\frac{d \alpha}{dt}
\biggl\vert_{t=t_{\alpha}} .
\eeq
From the last two equations, we get 
\beq
\frac{\d \bfX}{\d \alpha}={\bf 0}.
\eeq
From this equation, the function $\alpha=\alpha(t)$ is obtained.
 This is of the same form as the RG equation. Thus one may call the
 envelope equation the RG/E equation, too.
In the application of the envelope theory for constructing global
 solutions of differential equations, the parameter  is the initial time
 $t_0$, i.e., $\alpha =t_0$. 
 Actually, 
 apart from $t_0$, we have unknown functions given as initial values   
  in the applications. We use the above
 condition to determine the $t_0$ dependence of the initial values
 by imposing that $t_0=t$. In section 4, we shall show  that 
the resultant function obtained as the  envelope of the 
  local solutions in the  perturbation theory becomes an approximate but 
 uniformly valid solution.

\section{Envelope Surfaces}
\setcounter{equation}{0}
\renewcommand{\theequation}{\thesection.\arabic{equation}}

 This section is devoted to give the condition for constructing
 the envelope surface of a family of surfaces with two parameters
 in the three-dimensional space. The generalization to the 
$n$-dimensional case is straightforward.

Let $\{ {\rm S}_{\tau_1 \tau_2}\}_{\tau_1\tau_2}$ be a family of surfaces 
given by 
\beq
F({\bf r}; \tau_1, \tau_2)=0,
\eeq
and E the envelope surface of it given by
\beq
G({\bf r})=0,
\eeq
with ${\bf r}=(x, y, z)$.

The fact that E contacts with  S$_{\tau_1\tau_2}$ at $(x, y,z)$ implies
\beq
G({\bf r})=F({\bf r};\tau_1({\bf r}), \tau_2({\bf r}))=0.
\eeq
Let  $({\bf r}+d{\bf r}, \tau_1+d\tau_1, \tau_2+d\tau_2)$ gives another
point in E, then
\beq
G({\bf r}+d{\bf r})=F({\bf r}+d{\bf r};\tau_1+d\tau_1, \tau_2+d\tau_2)=0.
\eeq
Taking the difference of the two equations, we have
\beq
\nabla F\cdot d{\bf r}+\frac{\d F}{\d \tau_1}d\tau_1+
              \frac{\d F}{\d \tau_2}d\tau_2=0.
\eeq
On the other hand, the fact that E and S$_{\tau_1\tau_2}$ have a
common tangent plane at ${\bf r}$ implies that 
\beq
\nabla F\cdot d{\bf r}=0.
\eeq
Combining the last two equations, one has
\beq
\frac{\d F}{\d \tau_1}d\tau_1+\frac{\d F}{\d \tau_2}d\tau_2=0.
\eeq
Since $d\tau_1$ and $d\tau_2$ may be varied independently, we have
\beq
\frac{\d F}{\d \tau_1}=0,\ \ \ \ \frac{\d F}{\d \tau_2}=0.
\eeq
From these equations, we get $\tau_i$ as a function of ${\bf r}$;
$\tau _i=\tau_i({\bf r})$.

As an example, let
\beq
F(x, y, z; \tau_1, \tau_2)=\e ^{-\tau_1y}\{1-y(x-\tau_1)\}+\e ^{-\tau_2x}
\{1-x(y-\tau_2)\}-z.
\eeq
The conditions ${\d F}/{\d \tau_1}=0$ and ${\d F}/{\d \tau_2}=0$
 give 
\beq
\tau_1=x,\ \ \ \tau_2=y,
\eeq
 respectively. Hence one finds that 
the envelope is given by 
\beq
G(x, y, z)=F(x, y, z; \tau_1=x, \tau_2=y)=2\e ^{-xy}-z=0,
\eeq
or $z=2{\rm exp}(-xy)$.

It is obvious that the discussion can be extended to higher dimensional
 cases.
In Ref.\cite{kunihiro2}, envelope surfaces were constructed in multi 
steps when the RG method was applied to PDE's.  However, as 
 has been shown in this section, the construction can be performed by single
 step. 

\setcounter{section}{3}
\setcounter{equation}{0}
\section{The basis of the RG method for systems }
\renewcommand{\theequation}{\thesection.\arabic{equation}}

\subsection{ODE's}

Let ${\bf X}=\, ^t(X_1, X_2, \cdots , X_n)$ and 
 ${\bf F}({\bf X}, t; \eps) =\, ^t(F _1(\bfX, t; \eps)$, 
$F _2(\bfX, t; \eps),\cdots , F _n(\bfX, t; \eps))$, 
and $\bfX$ satisfy the equation
\beq
\dX = \bfF(\bfX , t; \eps).
\eeq
Let us try  to have the perturbation solution of 
Eq.(4.1) around $t=t_0$ by expanding 
\beq
\bfX (t; t_0)= \bfX _0(t; t_0) + \eps \bfX _1(t; t_0) 
               + \eps^2\bfX _2(t; t_0) \cdots.
\eeq
We suppose that an approximate solution 
$\tilde{\bfX}=\tbfX (t; t_0, {\bf W}(t_0))$ 
to the equation up to  $O(\eps ^p)$ is obtained,
\beq
\frac{d\tbfX  (t; t_0, {\bf W}(t_0))}{dt}=
\bfF (\tbfX (t), t; \eps) + O(\eps^p),
\eeq
where the $n$-dimensional vector 
${\bf W}(t_0)$ denotes the initial values assigned at the initial
 time $t=t_0$.  Here notice that $t_0$ is arbitrary.

Let us construct the envelope function $\bfX _E(t)$ of the family of 
trajectories given 
 by the functions $\tbfX(t; t_0, {\bf W}(t_0))$ with $t_0$  parameterizing the
trajectories. The construction is performed as follows: First we impose the
 RG/E equation, which now reads
\beq
\frac{d\tbfX}{d t_0}={\bf 0}.
\eeq
Notice that  $\tbfX$ contains the unknown function ${\bf W}(t_0)$ 
 of $t_0$.\footnote{ 
This means that Eq.(4.4) is a
 total derivative w.r.t. $t_0$;
\beq
\frac{d\tbfX}{d t_0}=\frac{\d\tbfX}{\d t_0}+ 
\frac{d{\bf W}}{d t_0}\cdot\frac{\d\tbfX}{\d {\bf W}}={\bf 0}.\nonumber 
\eeq
}
 In the usual theory of envelopes, as given in  section 2,
 this equation gives $t_0$ as a function
 of $t$. However, since we are now constructing the perturbation solution 
 that  is  as close as possible to the exact one around $t=t_0$, we demand 
that the RG/E equation should give the solution $t_0=t$, i.e.,
 the parameter 
should coincide with the point of tangency. It means that the RG/E equation
 should  determine the $n$-components of the initial
 vector ${\bf W}(t_0)$  so that $t_0=t$.  In fact,
Eq.(4.4) may  give equations as many as $n$ which are  independent 
 of each other.\footnote{In the applications given below, the equation
 is, however, reduced to a scalar equation.}
  Thus the envelope function is given by 
\beq
\bfX _E(t)=\tbfX ( t; t, {\bf W}(t)).
\eeq

Then the fundamental theorem for the RG method is the following:\\ 
{\bf Theorem:}\ \ {\em  $\bfX_E(t)$ satisfies the original
 equation uniformly up to $O(\eps ^p)$.}

{\bf Proof}  \ \  The proof is already given in Eq.(3.21)
 of Ref.\cite{kunihiro}.  Here we recapitulate it for completeness.
 $\forall t_0$, owing to the
 RG/E equation one has 
\beq
\frac{d\bfX_E}{dt}\Biggl\vert _{t=t_0} &=& 
\frac{d\tbfX(t; t_0, {\bf W}(t_0))}{d t}\Biggl\vert _{t=t_0}+
\frac{d\tbfX(t; t_0, {\bf W}(t_0))}{d t_0}\Biggl\vert _{t=t_0},
 \nonumber \\ 
 \ \ &=& \frac{d\tbfX(t; t_0, {\bf W}(t_0))}{d t}\Biggl\vert _{t=t_0},
\nonumber \\ 
 \ \ &=& \bfF (\bfX _E(t_0), t_0; \eps) + O(\eps^p),
\eeq
 where Eq.(4.4) has been used in the last equality.  This concludes the
 proof.

\subsection{PDE's}

It is desirable to develop a general theory for systems of PDE's as 
 has been done for ODE's.  But such a general theorem is not available 
yet.
 Nevertheless it {\em is} known that the simple generalization 
 of Eq. (4.4) to envelope surfaces works.
 
Let $\tbfX (t, \bfx : t_0, \bfx _0; \bfW (t_0, \bfx _0))$
is an approximate solution given in the perturbation theory up to 
 $O(\eps^p)$ of  
 a system of PDE's with respect to $t$ and $\bfx =(x_1, x_2, \dots , x_n)$.
Here we have made explicit that the solution has an initial and boundary 
 value $\bfW (t_0, \bfx _0)$ dependent on $t_0$ and 
$\bfx _0= (x_{10}, x_{20}, \dots , x_{n0})$.
As has been shown in section 3, the RG/E equation  now reads
\beq
\frac{d \tbfX}{d t_0}={\bf 0}, \ \ \ 
\frac{d \tbfX}{d x_{i0}}={\bf 0}, \ \ (i=1, 2, \dots , n).
\eeq
Notice again that $\tbfX$ contains the unknown function 
${\bf W}(t_0, {\bfx _0})$ dependent on $t_0$ and $\bfx _0$, hence
 the derivatives are total derivatives.
As the generalization of the case for ODE's, we demand that the RG/E
 equation should be compatible with the condition that the 
 coordinate of the point of tangency becomes the parameter of the
 family of the surfaces;i.e.,
\beq
t_0=t, \ \ \ \bfx _0 =\bfx.
\eeq
Then the RG/E equation is now reduced to equations for the 
unknown function $\bfW$, which will be shown to be
  the amplitude equations such as time-dependent Ginzburg-Landau
 equation. Here we remark that although Eq.(4.7) is a vector equation, 
 the equation to appear below will be  reduced to a 
scalar one; see subsection 6.2. 
It can be shown, at least  for equations treated so far and here, 
that the resultant envelope
 functions satisfy the original equations uniformly up to 
$O(\eps^p)$; see also Ref.\cite{kunihiro2}.

\section{Simple examples}
\setcounter{equation}{0}
\renewcommand{\theequation}{\thesection.\arabic{equation}}

In this section, we treat a few of simple examples of systems
of ODE's to show the how the RG method works. 
 The examples are the Duffing\cite{holmes} equation of non-autonomous 
nature, the  Lotka-Volterra\cite{lotka} and the Lorenz\cite{lorenz} 
equation.  The first one may be treated as
 a scalar equation.  Actually, the equation is easier to calculate when
 treated as a scalar one.  We give  such a treatment in Appendix B.
We shall work out to derive the time dependence  of the
 solution to the Lotka-Volterra equation explicitly.
 The last one is an example with three degrees of freedom, which shows 
a  bifurcation\cite{holmes}.  We shall give the center manifolds to this
 equation around the first bifurcation of the Lorenz model.
A general treatment for
 equations with a bifurcation will be treated in section 6.

\subsection{Forced Duffing equation}

 The  forced Duffing equations are reduced to 
\beq
\ddot {x}+ 2\eps \gamma \dot{x}+ (1+\eps \sigma)x + \eps hx^3&=&
\eps f\cos t, \nonumber \\
\ddot {y}+ 2\eps \gamma \dot{y}+ (1+\eps \sigma)y + \eps hy^3 &=&
\eps f\sin t.
\eeq
Defining a complex variable $z=x+i y$, one has
\beq
\ddot {z}+ 2\eps \gamma \dot{z}+ (1+\eps \sigma)z +
\frac{\eps h}{2}(3\vert z\vert^2z +{z^{\ast}}^3)= \eps f\e^{it}.
\eeq
We suppose that $\eps$ is small.

We convert the equation to the system 
\beq
\biggl(\frac{d}{dt} -L_0\biggl)\bfu = -\eps F(\xi, \eta; t)
                                      \pmatrix{0\cr 1},
\eeq
where
\beq
\bfu &=& \pmatrix{\xi \cr \eta}, \ \ \ 
\xi= z, \ \ \eta = \dot{z},\nonumber \\
L_0 &=& \pmatrix{\ 0 & 1\cr
               -1& 0},
\eeq
and
\beq
F(\xi, \eta; t)=\sigma \xi + 2\gamma \eta \frac{h}{2}(3\vert \xi\vert^2
   + {\xi ^{\ast}}^3) - f\e^{it}.
\eeq

Let us first solve the equation in the perturbation theory by expanding
\beq
\bfu = \bfu _0 + \eps \bfu _1 + \dots,
\eeq
with $\bfu _i=\ ^t(\xi _i, \eta  _i)$\, $(i=0, 1, \dots)$.
We only have to solve the following equations successively;
\beq
\biggl(\frac{d}{dt} -L_0\biggl)\bfu _0&=&{\bf 0}, \nonumber \\
\biggl(\frac{d}{dt} -L_0\biggl)\bfu _1&=&
 - F(\xi _0, \eta _0; t)\pmatrix{0\cr 1},
\eeq
and so on.
The solution of the zero-th order equation is  found to be
\beq
\bfu _0(t; t_0)= W(t_0)\bfU \e^{it},
\eeq
where $\bfU $ is an eigenvector belonging to an eigen value $i$ of $L_0$,
\beq
L_0\bfU = i\bfU, \ \ \ \bfU=\pmatrix{1\cr i}.
\eeq
The other eigenvector is given by the complex conjugate $\bfU^{\ast}$,
 which belongs to the other eigenvalue $-i$.  We have made it explicit 
that the constant $W$ may be dependent on the initial time $t_0$.
In terms of the component,
\beq
\xi_0(t;t_0)= W(t_0)\e ^{it}, \ \ \ 
\eta _0(t;t_0) = iW(t_0)\e^{it}.
\eeq

Inserting these into $F(\xi _0, \eta _0;t)$, one has
\beq
F(\xi _0, \eta _0;t)={\cal W}(t_0)\e^{it} +
  \frac{h}{2}{W^{\ast}}^3\e^{-3it},
\eeq
with
\beq
{\cal W}(t_0)\equiv (\sigma +2i\gamma)W + \frac{3h}{2}\vert W\vert ^2W -f
\eeq
We remark  that the inhomogeneous term includes a term proportional to 
 the zero-th order solution. Thus  $\bfu _1$ contains a resonance
 or a secular term as follows;
\beq
\bfu _1(t; t_0)&=& -\frac{1}{2i}{\cal W}\e^{it}\{(t-t_0 +\frac{1}{2i})
     \bfU -\frac{1}{2i}\bfU ^{\ast}\} 
  -\frac{h}{16}{W^{\ast}}^3\e^{-3it}(\bfU -2\bfU ^{\ast}).
\eeq
In terms of the components
\beq
\xi _1(t;t_0)&=& \frac{i}{2}{\cal W}\e ^{it}(t-t_0)+\frac{h}{16}{W^{\ast}}^3
 \e^{-3it}, \nonumber \\
\eta _1(t; t_0)&=& -\frac{{\cal W}}{2}\e^{it}(t-t_0 - i)
    -\frac{3i}{16}h{W^{\ast}}^3\e^{-3it}.
\eeq

Adding the terms, we have
\beq
\bfu(t)&\simeq& \bfu_0(t;t_0) + \eps \bfu_1(t;t_0), \nonumber \\ 
 \ \ \ &=& W(t_0)\bfU \e^{it}-
   \eps \frac{1}{2i}{\cal W}\e^{it}\{(t-t_0 +\frac{1}{2i})
     \bfU -\frac{1}{2i}\bfU ^{\ast}\}
  -\eps \frac{h}{16}{W^{\ast}}^3\e^{-3it}(\bfU -2\bfU ^{\ast}),
   \nonumber \\ 
 \ \ \ &\equiv& \tilde{\bfu}(t;t_0).
\eeq
In terms of the components,
\beq
\xi(t;t_0)&\simeq&W(t_0)\e^{it} +\eps \frac{i}{2}{\cal W}(t_0)\e^{it}(t-t_0)
  +\eps \frac{h}{16}{W^{\ast}}^3\e^{-3it}\equiv \tilde{\xi},\nonumber \\
\eta(t;t_0)&\simeq&iW(t_0)\e^{it}-\eps\frac{{\cal W}}{2}\e^{it}(t-t_0 -i)
 -\eps \frac{3i}{16}h{W^{\ast}}^3\e^{-3it}\equiv\tilde{\eta}.
\eeq

Now let us construct the envelope $\bfu_E(t)$ of the family of 
trajectories or curves 
 $\tilde{\bfu}(t; t_0)=(\tilde{\xi}(t;t_0), \tilde{\eta}(t;t_0))$
 which is parametrized with $t_0$; $\bfu_E(t)$ will be found to be an
 approximate solution to Eq. (5.3) in the global domain.  According 
 to section 2, the envelope
 may be obtained from the equation
\beq
\frac{d\tilde{\bfu}(t;t_0)}{d t_0}=0.
\eeq
In the usual procedure for constructing the envelopes, the above equation
is used for obtaining $t_0$ as a function of $t$, and the resulting
 $t_0=t_0(t)$ is inserted in $\tilde{\bfu}(t;t_0)$ to make the 
envelope function $\bfu _E(t)=\tilde{\bfu}(t; t_0(t))$.  In our case,
 we are constructing the envelope around $t=t_0$, so we rather impose
 that 
\beq
t_0=t,
\eeq
and Eq.(5.17) is used to obtain the initial value
 $W(t_0)$ as a function of $t_0$. That is, we have
\beq
0&=&\frac{d\tilde{\bfu}(t;t_0)}{d t_0}\biggl\vert _{t_0=t},\nonumber \\
 \ &=& 
  \frac{dW}{dt}\bfU \e^{it} +\eps\frac{{\cal W}}{2i}\e^{it}\bfU +
\eps\frac{i}{2}\frac{d{\cal W}}{dt}\e^{it}\frac{1}{2i}(\bfU -\bfU ^{\ast})
 -\frac{3\eps h}{16}\frac{dW^{\ast}}{dt}\e^{-3it}(\bfU -2\bfU ^{\ast}).
\eeq
Noting that the equation is consistent with  $dW/dt=O(\eps)$,  one has 
\beq
\frac{dW}{dt}&=& i\frac{\eps}{2}{\cal W}(t),\nonumber \\ 
\ \ \ &= & i\frac{\eps}{2}\{(\sigma +2i\gamma)W(t)+ \frac{3h}{2}
\vert W(t)\vert ^2 W(t) -f\}.
\eeq
This is the amplitude equation called Landau-Stuart equation, 
which may be also given by the 
 RP method\cite{kuramoto} as a reduction of
 the dynamics.
With this equation, the envelope trajectory is given by 
\beq
\xi_E(t)&=& W(t)\e^{it} + \eps \frac{h}{16}{W^{\ast}}^3\e^{-3it},
\nonumber \\ 
\eta _E(t)&=& i(W(t)+\eps \frac{1}{2}{\cal W}(t))\e^{it}
  -\eps \frac{3i}{16}h{W^{\ast}}^3\e^{-3it}.
\eeq

For completeness, let us examine the stationary solution of the
 Landau-Stuart equation, briefly;
\beq
{\cal W}=(\sigma +2i\gamma)W + \frac{3}{2}\eps h\vert W\vert ^2W-f=0.
\eeq
Writing $W$ as 
\beq
W=A\e ^{-i\theta},
\eeq
we have
\beq
A^2\biggl[(\frac{3}{2}hA^2+\sigma)^2+4\gamma^2\biggl]=f^2,
\eeq
which describes the jumping phenomena of the Duffing oscillator.

\subsection{Lotka-Volterra equation}
     
As another simple  example, we take 
the Lotka-Volterra equation\cite{lotka};
\beq
\dot{x}= ax -\eps xy, \ \ \ \ \dot{y}=-by+\eps'xy,
\eeq
where the constants $a, b, \eps$ and $\eps'$ are assumed to be positive.
It is well known that the equation has the  conserved quantity, i.e., 
\beq
b\ln\vert x\vert + a\ln \vert y\vert -(\eps' x+\eps y)={\rm const.}.
\eeq 

The fixed points are given by $(x=0, y=0)$ and $(x=b/\eps', y=a/\eps)$.
Shifting and scaling  the variables by
\beq
x=(b+ \eps\xi)/\eps', \ \ \ \ y=a/\eps + \eta,
\eeq
we get the reduced equation given by the system
\beq
\biggl(\frac{d}{dt}- L_0\biggl)\bfu= -\eps\xi\eta\pmatrix{\ 1\cr -1},
\ \ \ \ 
\eeq
where  
\beq
\bfu = \pmatrix{\xi\cr \eta},\ \ \ \ L_0=\pmatrix{0 & -b\cr a & \ 0}.
\eeq
The eigen value equation
\beq
L_0\bfU=\lambda _0\bfU
\eeq
has the solution
\beq
\lambda _0=\pm i\sqrt{ab}\equiv \pm i\omega, \ \ \ \ 
\bfU =\pmatrix{\, 1\cr \mp i\frac{\omega}{b}}.
\eeq

Let us try to apply the perturbation theory to solve the equation
by  expanding the variable in a Taylor series of $\eps$;
\beq
\bfu=\bfu_0+\eps\bfu_1 +\eps^2\bfu_2+\cdots,
\eeq
with $\bfu _i=\ ^t(\xi _i, \eta_i)$.
The lowest term satisfies the equation
\beq
\biggl(\frac{d}{dt}- L_0\biggl){\bfu}_0={\bf 0},
\eeq
which yields the solution
\beq
\bfu _0(t;t_0)=W(t_0){\e}^{i\omega t}\bfU + {\rm c.c.},
\eeq
or
\beq
\xi _0= W(t_0)\e ^{i\omega t} + {\rm c.c.}, \ \ \ \ 
\eta _0=-\frac{\omega}{b}\big(iW(t_0)\e ^{i\omega t} + {\rm c.c.}\big).
\eeq
 Here we have supposed that the initial value
 $W$ depends on the initial time $t_0$.

Noting that 
\beq
\pmatrix{\ 1\cr -1}=\alpha \bfU + {\rm c.c.},
\eeq
with $\alpha=(1- ib/\omega)/2$, one finds that 
the first order term  satisfies the equation
\beq
\biggl(\frac{d}{dt} - L_0\biggl)\bfu _1=
 \frac{\omega}{b}\biggl[iW^2 \e ^{2i\omega t}
  (\alpha \bfU + {\rm c.c.}) + {\rm c. c.}\biggl],
\eeq
  the solution to which is found to be
\beq
\bfu _1=\frac{1}{b}\biggl[W^2(\alpha \bfU + \frac{\alpha ^{\ast}}{3}
\bfU ^{\ast})
 \e^{2i\omega t} + {\rm c.c.}\biggl],
\eeq
or
\beq
\xi _1 &=&\frac{1}{b}\bigl( \frac{2\omega - ib}{3\omega}W^2\e ^{2i\omega t}
 + {\rm c.c.}\bigl), \nonumber \\ 
\eta _1 &=& -\frac{\omega}{3b^2}
\bigl( \frac{2b+ i\omega }{\omega}W^2\e ^{2i\omega t}
 +{\rm c.c.}\bigl).
\eeq

The second  order equation now reads
\beq
\biggl(\frac{d}{dt} - L_0\biggl)\bfu _2 = 
 \frac{1}{3b^2}\biggl[\{ (b-i\omega)\vert W\vert ^2W\e^{i\omega t}
 + 3(b+i\omega)W^3\e ^{3i\omega t}\} + {\rm c.c.}\biggl]\pmatrix{\ 1\cr -1}.
\eeq
 We remark that the inhomogeneous term has a part  proportional to the
 zero-th-order solution, which gives rise to a resonance. Hence the  
 solution necessarily includes secular terms as follows;
\beq
\bfu _2&=& \Biggl[\frac{b-i\omega}{3b^2}\vert W\vert ^2W
\biggl\{ \alpha (t-t_0 +i\frac{\alpha^{\ast}}{2\omega})
         \bfU + \frac{\alpha ^{\ast}}{2i\omega}\bfU ^{\ast}\biggl\}
           \e ^{i\omega t } \nonumber \\ 
 \ \ \  & & + \frac{b+i\omega}{4b^2i\omega}W^3(2\alpha \bfU +
 \alpha^{\ast}\bfU ^{\ast})\e^{3i\omega t}\Biggl]
+ {\rm c.c.} .
\eeq
In terms of the components, one finds
\beq
\xi _2 &=& 
\Biggl[ \frac{-i}{6\omega}\frac{b^2+\omega^2}{b^2}\vert W\vert ^2W(t-t_0)
 \e ^{i\omega t}\nonumber + 
  \frac{W^3}{8b^2\omega ^2}\{ (3\omega ^2 -b^2) 
- 4ib\omega \}\e ^{3i\omega t}\Biggl] + {\rm c.c.}  \nonumber \\ 
\eta _2 &=& \frac{\vert W\vert ^2W}{6b^3}
\Biggl[
-(b^2 +\omega^2)(t-t_0) +\frac{1}{\omega}
\{2b\omega +i (b^2 -\omega ^2)\}\Biggl]\e^{i\omega t}\nonumber \\ 
  \ \ \ \ & \ & 
+ \frac{W^3}{8b^3}\{ -4b + \frac{i}{\omega}(3b^2 -\omega ^2)\}
\e^{3i\omega t}
             + {\rm c.c.} .
\eeq

The RG/E equation reads
\beq
\frac{d \bfu}{d t_0}={\bf 0},
\eeq
 with $t_0=t$,  which gives the equation for $W(t)$ as
\beq
\frac{d W}{dt}= - i\eps^2 \frac{\omega ^2+b^2}{6\omega b^2}\vert W\vert ^2 W.
\eeq 
If we define $A(t)$ and $\theta (t)$ by 
$W(t)=(A(t)/2i) {\rm exp} i\theta(t)$, the equation gives
\beq
A(t)= {\rm const.}, \ \ \ \ 
\theta (t) = - \frac{\eps^2A^2}{24}(1+ \frac{b^2}{\omega ^2})\omega t
  + \bar{\theta }_0,
\eeq
with $\bar{\theta }_0$ being a constant.  Owing to the  prefactor $i$ 
in r.h.s. of Eq. (5.44), the absolute value of the amplitude $A$ becomes
 independent of $t$, while the phase $\theta$ has a $t$-dependence.
  The envelope function is given by
\beq
\bfu _E(t)=\pmatrix{\xi _E(t)\cr \eta _E(t)}=
\bfu (t, t_0)\Biggl\vert_{t_0=t, \d \bfu/\d t_0=0}.
\eeq
In terms of the components, one has
\beq
\xi _{_E}&= & A\sin \Theta (t) -
\eps \frac{A^2}{6\omega}(\sin 2\Theta (t)
 + \frac{2\omega }{b}\cos 2\Theta (t))\nonumber \\ 
 \ \ \ & \ & -\frac{\eps^2 A^3}{32}\frac{3\omega ^2 -b^2}{\omega ^2b^2}
(\sin 3\Theta (t) - \frac{4\omega b}{3\omega ^2 -b^2}\cos 3\Theta (t) ),
\nonumber \\ 
\eta _{_E} &=& -\frac{\omega}{b}\Biggl[
 \biggl(A - \frac{\eps^2A^3}{24}\frac{b^2-\omega ^2}{b^2\omega ^2}\biggl)
\cos \Theta (t) - \frac{\eps ^2 A^3}{12b\omega}\sin \Theta (t)
\nonumber \\ 
 \ \ \ \ & \ & + \eps \frac{A^2}{2b}\biggl(\sin 2\Theta (t) - 
\frac{2b}{3\omega}\cos 2\Theta (t)\biggl)
  - \frac{\eps^2A^3}{8b\omega}\biggl( \sin 3\Theta (t)
- \frac{3b^2 -\omega ^2}{4b^2\omega ^2}\cos 3\Theta (t)\biggl)\Biggl],
\eeq
where
\beq
\Theta (t) \equiv \tilde {\omega} t + \bar{\theta}_0, 
\ \ \ \ \tilde {\omega} \equiv \{ 
1- \frac{\eps^2A^2}{24}(1+ \frac{b^2}{\omega ^2})\}\omega .
\eeq
One sees that the angular frequency is shifted.

 We identify $\bfu_E(t)= (\xi _E(t), \eta _E(t))$ as an approximate 
 solution to Eq.(5.28). 
According to the basic theorem presented in section 4, $\bfu_E(t)$
is an approximate but uniformly valid solution to the equation
 up to $O(\eps^3)$.  We remark that 
 the resultant trajectory is closed in conformity  with the conservation law 
given in Eq. (5.26).

``Explicit solutions'' of two-pieces of Lotka-Volterra equation were
 considered by Frame \cite{frame}; however, his main conceren was on 
 extracting the  period of the solutions in an average method.
 Comparing the Frame's method,
 the RG method is simpler, more transparent and explicit.
 The present author is not aware of any other 
work which gives an explicit form of the solution as given in Eq. (5.47,48).

\subsection{The Lorenz model}

The Lorenz model\cite{lorenz} for the thermal convection is given by 
\beq
\dot{\xi}&=&\sigma(-\xi+\eta),\nonumber \\
\dot{\eta}&=& r\xi -\eta -\xi\zeta,\nonumber \\
\dot{\zeta}&=& \xi\eta - b \zeta.
\eeq
The steady states are give by 
\beq
{\rm (A)}\ \ (\xi, \eta, \zeta)=(0, 0, 0),\ \ \ 
{\rm (B)}\ \ (\xi, \eta, \zeta)=
(\pm \sqrt{b(r-1)},\pm \sqrt{b(r-1)},r-1).
\eeq

The linear stability analysis\cite{holmes} shows that the origin is stable for 
$0<r<1$ but unstable for $r>1$, while the latter steady states (B) are 
 stable for $1<r<\sigma(\sigma+b+3)/(\sigma -b-1)\equiv r_c$ but 
unstable for $r>r_c$.
In this paper, we examine the non-linear stability around the origin
 for $r\sim 1$; we put 
\beq
r=1+\mu \ \ \ {\rm and}\ \ \ 
\mu =\chi \eps^2, \ \ \ \chi={\rm sgn}\mu.
\eeq
We expand the quantities as Taylor series of $\eps$:
\beq
\bfu\equiv \pmatrix{\xi\cr 
                   \eta\cr 
                   \zeta}
 = \eps \bfu_1+\eps^2\bfu_2 + \eps ^3\bfu_3 + \cdots,
\eeq
where $\bfu _i=\ ^t(\xi_i, \eta_i, \zeta_i) $\ $(i=1, 2, 3, \dots)$.
The first order equation reads
\beq
\biggl(\frac{d}{dt} - L_0\biggl)\bfu_1={\bf 0},
\eeq
where
\beq
L_0=\pmatrix{-\sigma & \sigma & 0\cr 
                 1   &    -1  & 0\cr
                 0   &    0   & -b},
\eeq
the eigenvalues of which are found to be
\beq
\lambda _1=0, \ \ \ \lambda _2= - \sigma -1,\ \ \ \lambda _3= -b.
\eeq
The respective eigenvectors are 
\beq
\bfU _1=\pmatrix{1\cr
                 1\cr
                 0}, \ \ \ 
\bfU _2=\pmatrix{\sigma\cr
                 -1\cr
                 0}, \ \ \ 
\bfU _3=\pmatrix{0\cr
                 0\cr
                 1}.
\eeq
When we are interested in the asymptotic state as $t\rightarrow \infty$, 
one may take the neutrally stable solution
\beq
\bfu _1(t; t_0)=W(t_0)\bfU_1,
\eeq
where we have made it explicit that the solution may depend on the
 initial time $t_0$,  which is supposed to be close to $t$.
In terms of the components,
\beq
\xi_1(t)=W(t_0), \ \ \ \eta_1(t)=W(t_0), \ \ \ \zeta _1(t) =0.
\eeq

The second order equation now reads
\beq
\biggl(\frac{d}{dt} - L_0\biggl)\bfu_2=\pmatrix{\ \ 0\cr
                                                -\xi_1\zeta_1\cr
                                                \xi_1\eta_1}
                                     = W^2\bfU_3,
\eeq
which yields
\beq
\bfu_2(t)=\frac{W^2}{b}\bfU_3,
\eeq
 or in terms of the components
\beq
\xi_2=\eta_2=0, \ \ \  \zeta_2=\frac{W^2}{b}.
\eeq
Then the third order equation is given by
\beq
\biggl(\frac{d}{dt} - L_0\biggl)\bfu_3= 
      \pmatrix{\ \ \ 0\cr
               -\chi\xi_1-\xi_2\zeta_1-\xi_1\zeta_2\cr
               \xi_2\eta_1+\xi_1\eta_2}
    = \frac{1}{1+\sigma}(\chi W-\frac{1}{b}W^3)(\sigma\bfU_1 -\bfU_2),
\eeq
which yields
\beq
\bfu_3=\frac{1}{1+\sigma}(\chi W-\frac{1}{b}W^3)
      \{\sigma(t-t_0 + \frac{1}{1+\sigma})\bfU_1 -
   \frac{1}{1+\sigma}\bfU_2\}.
\eeq

Thus gathering all the terms, one has 
\beq
\bfu (t;t_0)&=& 
\eps W(t_0)\bfU_1 + \frac{\eps^2}{b}W(t_0)^2\bfU_3 \nonumber \\ 
 \ \ \ \ &\ & \ \ \ 
 + \frac{\eps ^3}{1+\sigma}(\chi W(t_0) -\frac{1}{b}W(t_0)^3)
      \{\sigma(t-t_0 + \frac{1}{1+\sigma})\bfU_1 -
   \frac{1}{1+\sigma}\bfU_2\},
\eeq
 up to $O(\eps ^4)$.
The RG/E equation now reads
\beq
{\bf 0}&=&\frac{d \bfu}{d t_0}\biggl\vert_{t_0=t},\nonumber \\ 
\ &=& \eps \frac{dW}{dt}\bfU_1+ 2 \frac{\eps^2}{b}W\frac{dW}{dt}\bfU_3
 -\frac{\sigma}{1+\sigma}\eps^3(\chi W - \frac{1}{b}W^3)\bfU_1,
\eeq 
up to $O(\eps^4)$. Noting that one may self-consistently assume that 
 $dW/dt=O(\eps^2)$,  we have the amplitude equation
\beq
\frac{dW}{dt}=\eps^2\frac{\sigma}{1+\sigma}(\chi W(t) - \frac{1}{b}W(t)^3).
\eeq
With this $W(t)$, the envelope function is given by 
\beq
\bfu_E(t)&=&\bfu (t; t_0=t),\nonumber \\ 
 \ \ \ &=& 
\eps W(t)\bfU_1 + \frac{\eps^2}{b}W(t)^2\bfU_3 
+ \frac{\eps ^3}{(1+\sigma)^2}(\chi W(t) -\frac{1}{b}W(t)^3)
      (\sigma \bfU_1 -\bfU_2),
\eeq
or
\beq
\xi_E(t)&=&\eps W(t),\nonumber \\ 
\eta_E(t)&=& \eps W(t) +\frac{\eps^3}{1+\sigma}
           (\chi W(t)-\frac{1}{b}W(t)^3),\nonumber \\ 
\zeta_E(t)&=& \frac{\eps^2}{b}W(t)^2.
\eeq
We may identify the envelope functions thus constructed as a global
 solution to the Lorenz model; according to the general theorem
  given in section 4, 
the envelope functions satisfy Eq.(5.49) approximately but 
 uniformly for $\forall t$ up to $O(\eps ^4)$.

A remark is in order here; Eq.(5.68) shows that the slow manifold which 
 may be identified with a center manifold\cite{holmes} is given by 
\beq
\eta=(1+ \eps^2\frac{\chi}{1+\sigma})\xi - \frac{1}{b(1+\sigma)}\xi^3,
 \ \ \ \zeta= \frac{1}{b}\xi^2.
\eeq
Notice here that the RG method is also a powefull tool to extract center 
manifolds in a concrete form. It is worth mentioning that since the 
RG method utilizes
 neutrally stable solutions as the unperturbed ones, it is rather natural
 that the RG method can extract center manifolds when exist.
 The applicability of the RG method was discussed in
 \cite{goldenfeld2} using a generic model having a center manifold,
 although the relation between the exitence of  center manifolds and 
 neutrally stable solutions is not so transparent  in their general 
 approach.

\setcounter{equation}{0}
\section{Bifurcation Theory}
\renewcommand{\theequation}{\thesection.\arabic{equation}}
 
In this section, we  take   generic equations 
 with a bifurcation. We shall derive the Landau-Stuart and Ginzburg-Landau
 equations in the RG method.
  In this section, we shall follow Kuramoto's monograph\cite{kuramoto}
 for notations to clarify the correspondence between the RG method and
 the reductive perturbation (RP) method. 

\subsection{Landau-Stuart equation}

We start with the $n$-dimensional equation
\beq
\dX = \bfF(\bfX , t; \mu ).
\eeq
Let ${\bf X}_0(\mu)$ is a steady solution 
\beq
\bfF({\bf X}_0(\mu) ; \mu)=0.
\eeq
Shifting the variable as $\bfX = {\bf X}_0 + {\bf u}$,
we have a Taylor series
\beq
\du = L\bfu + M\bfu \bfu + N\bfu \bfu \bfu+\cdots ,
\eeq
where we have used the diadic and triadic notations\cite{kuramoto};
\beq
L_{ij}&=&\frac{\d F_i}{\d X_j}\biggl\vert _{\bfX =\bfX _0}, \ \ \ 
(M\bfu \bfu)_i=\sum _{j, k}
\half \frac{\d ^2 F_i}{\d X_j\d X_k}\biggl\vert _{\bfX =\bfX _0}u_ju_k,
\nonumber \\ 
(N\bfu \bfu \bfu)_i&=&  \sum _{j, k, l}
\frac{1}{6}\frac{\d ^3 F_i}{\d X_j\d X_k\d X_l}
\biggl\vert _{\bfX =\bfX _0}u_ju_ku_l.
\eeq

We suppose that when $\mu<0$, ${\bf X}_0$ is stable for sufficiently
 small perturbations, while when $\mu >0$, otherwise. We also 
 confine ourselves to the case where a Hopf bifurcation occurs.
We expand  $L, M$ and $N$ as 
\beq
L=L_0 + \mu L_1 + \cdots , \ \ M=M_0 + \mu M_1 + \cdots ,
\ \ N=N_0 + \mu N_1 + \cdots . 
\eeq
The eigenvalues $\lambda^{\alpha}\, (\alpha=1, 2, \dots , n)$ of $L$
 are also expanded as
\beq
\lambda^{\alpha}=\lambda^{\alpha}_0 + \mu \lambda^{\alpha}_1 + \cdots,
\eeq
with
\beq
L_0\bfU _{\alpha}= \lambda^{\alpha}_0\bfU _{\alpha}.
\eeq 
We assume that  $\lambda^{1}_0=-\lambda^{2}_0$ are
 pure imaginary, i.e., $\lambda^{1} _0 = i\omega_0$, and 
 $\Re \lambda^{\alpha}_0<0$ for $\alpha=3, 4, \dots$.

Defining $\eps$ and $\chi$ by $\eps = \sqrt{\vert \mu\vert}$
 and $\chi={\rm sgn}\mu$, we expand as
\beq
\bfu = \eps \bfu _1 + \eps^2 \bfu_2 + \eps ^3\bfu_3 +\cdots.
\eeq
The $\bfu _i$ $(i=1, 2, 3, ...)$ satisfies 
\beq
\biggl(\frac{d}{dt} - L_0\biggl)\bfu _1&=& {\bf 0}, \nonumber \\
\biggl(\frac{d}{dt} - L_0\biggl)\bfu _2&=& M_0\bfu _1\bfu _1, \nonumber \\
\biggl(\frac{d}{dt} - L_0\biggl)\bfu _3&=& \chi L_1\bfu _1 + 2M_0\bfu _1
\bfu _2 + N_0 \bfu _1\bfu _1 \bfu _1,
\eeq
etc.

To see the asymptotic behavior as $t\rightarrow \infty$, 
  we take the neutrally stable solution as the lowest one  around
 $t= t_0$;
\beq
\bfu _1 (t; t_0)=W(t_0)\bfU {\rm e}^{i\omega_0t} + {\rm c.c.},
\eeq
where  c.c. stands for the complex conjugate.  With this choice, we have
 only two degrees of freedom for the initial value $W(t_0)$.

The second order equation is solved easily to yield
\beq
\bfu _2(t;t_0)= \bigl({\bf V}_{+}W(t_0)^2 {\rm e}^{2i\omega_0 t} + 
 {\rm c. c.} \bigl) + 
         {\bf V}_0\vert W(t_0)\vert ^2,
\eeq
where
\beq
{\bf V}_{+}= - (L_0-2i\omega _0)^{-1} M_0\bfU \bfU,\ \ \ 
{\bf V}_{0}= - 2L_0^{-1} M_0\bfU \bar{\bfU},
\eeq
with $\bar{\bfU}$ being the complex conjugate of $\bfU$.\footnote{
 In other sections, we use the notation $a^{\ast}$ for the 
complex conjugate of $a$.  In this section, $^{\ast}$ is used for a 
 different meaning, following ref.\cite{kuramoto}; see Eq. (6.16).}
Inserting $\bfu _1$ and $\bfu _2$ into the r.h.s of Eq. (6.9),
 we get 
\beq
\biggl(\frac{d}{dt} - L_0\biggl)\bfu _3 &=&\bigl\{\chi L_1 W\bfU + 
      (2M_0\bar{\bfU}{\bf V}_{+} + 3N_0\bfU \bfU\bar{\bfU})
\vert W\vert ^2W\bigl\}{\rm e}^{i\omega_0t} + {\rm c.c.} + {\rm h.h.}, 
  \nonumber \\ 
 \ \ \ & \equiv & {\bf A}{\rm e}^{i\omega_0t}
  + {\rm c.c.} + {\rm h.h.},
\eeq 
where h.h. stands for  higher harmonics.
 So far, the discussion is a simple perturbation theory
 and has proceeded in the same way 
as given in the RP method except for not having introduced  multiple times.

Now we expand  ${\bf A}$  by the eigenvectors 
 $\bfU _{\alpha}$ of $L_0$ as 
\beq
{\bf A}=\sum _{\alpha}A_{\alpha}\bfU _{\alpha},
\eeq
where 
\beq
A_{\alpha}= \bfU ^{\ast}_{\alpha}{\bf A}.
\eeq
Here $\bfU ^{\ast}_{\alpha}$ satisfies 
\beq
\bfU ^{\ast}_{\alpha}L_0=\lambda^{\alpha}_0L_0 ,
\eeq
and 
 is normalized as $\bfU^{\ast}_{\alpha}\bfU_{\alpha} =1$. 

 Then we get for $\bfu_3$
\beq
\bfu_3(t;t_0)=\{A_1(t-t_0+\delta)\bfU + 
\sum _{\alpha\not= 1}\frac{A_{\alpha}}{i\omega_0 - \lambda_0^{\alpha}}
\bfU_{\alpha}\}{\rm e}^{i\omega_0t}
 + {\rm c.c.} + {\rm h.h.}.
\eeq
The constant $\delta$ is chosen so that the coefficient of the
 secular term  of the first component vanishes at $t=t_0$.
 Note the appearance of the secular term which was 
 to be avoided in the RP method: The condition
 for the secular terms to vanish is called the solvability condition 
 which plays the central role in the RP method\cite{kuramoto}.

Thus we finally get
\beq
\bfu(t;t_0)=\{\eps W(t_0)\bfU 
  + \eps ^3 \bigl(A_1(t-t_0+ \delta)\bfU + 
\sum _{\alpha\not= 1}\frac{A_{\alpha}}{i\omega_0 - \lambda_0^{\alpha}}
\bfU_{\alpha}\bigl)\}{\rm e}^{i\omega_0t}+ {\rm c.c.} + {\rm h.h.}.
 \eeq

 The RG/E equation 
\beq
\frac{d {\bfu}}{d t_0}\Biggl\vert_{t_0=t}={\bf 0},
\eeq
 yields
\beq
\frac{dW}{dt}&=&\eps ^2A_1, \nonumber \\ 
\ \ &=& \eps^2\bigl[
  \chi \bfU ^{\ast}L_1\bfU W+ \{ 2\bfU^{\ast}M_0\bar{\bfU}{\bf V}_{+}
 +3\bfU^{\ast}N_0\bfU\bfU\bar{\bfU}\}\vert W\vert^2W\bigl] , 
\eeq
up to $O(\eps^3)$.  Here note that the terms coming from h.h. 
do not contribute  to this order because 
 $dW/dt_0$ is $O(\eps ^2)$.
The resultant equation is so called the Landau-Stuart equation and
 coincides with the result derived in the RP method\cite{kuramoto}.

\subsection{The Ginzburg-Landau equation}

We add the diffusion term to Eq.(6.1);
\beq
\dX = \bfF(\bfX )+ D\nabla ^2 \bfX,
\eeq
where $D$ is a diagonal matrix.  Let $\bfX _0$ be a uniform and steady 
solution.

Shifting the variable $\bfX = \bfX _0 +\bfu$ as before, we have
\beq
\du = \hat{L}\bfu + M\bfu \bfu + N\bfu \bfu \bfu+\cdots ,
\eeq
with
\beq
\hat{L} = L +D\nabla ^2.
\eeq

Then using the same expansion as before, we have the same equation for
$\bfu _1,  \bfu_2$ and $\bfu_3$ as given in Eq.(6.9) with $L_0$
 being replaced with $\hat{L}_0\equiv L_0 + D\nabla ^2$.

To see the asymptotic behavior as $t\rightarrow \infty$, 
  we take the neutrally stable uniform solution as the lowest one  around
 $t= t_0$ and ${\bf r}={\bf r}_0$;
\beq
\bfu _1 (t, {\bf r}; t_0, {\bf r}_0) = 
W(t_0, {\bf r}_0)\bfU {\rm e}^{i\omega_0t} + {\rm c.c.}.
\eeq
 With this choice, we have
 only two degrees of freedom for the initial value $W(t_0, {\bf r}_0)$.

The second order equation is solved easily to yield  the same form as
 that given in Eq.(6.11).
Inserting $\bfu _1$ and $\bfu _2$ into the r.h.s of Eq. (6.9) with 
$L_0$ replaced with $\hat{L}_0$,
 we have
\beq
\biggl(\frac{\d}{\d t} - \hat{L}_0\biggl)\bfu _3 &=&\bigl\{\chi L_1 W\bfU + 
      (2M_0\bar{\bfU}{\bf V}_{+} + 3N_0\bfU \bfU\bar{\bfU})
\vert W\vert ^2W\bigl\}{\rm e}^{i\omega_0t} + {\rm c.c.} + {\rm h.h.}, 
  \nonumber \\ 
 \ \ \ & \equiv & {\bf A}{\rm e}^{i\omega_0t}
  + {\rm c.c.} + {\rm h.h.} .
\eeq 
 Then we get for $\bfu_3$ in the spatially 1-dimensional case,
\beq
\bfu_3(t;t_0)&=&\biggl[A_1\{c_1(t-t_0+\delta)
-\frac{c_2}{2} D^{-1}(x^2 -x_0^2+\delta')\}\bfU + 
\sum _{\alpha\not= 1}\frac{A_{\alpha}}{i\omega_0 - \lambda_0^{\alpha}}
\bfU_{\alpha}\biggl]{\rm e}^{i\omega_0t} \nonumber \\
 \ \ \  &\ & + {\rm c.c.} + {\rm h.h.},
\eeq
with $c_1+c_2=1$.
 We have introduced constants $\delta$ and $\delta'$ so that the 
 secular terms of the first component of $\bfu _3$ vanish at $t=t_0$
 and $x=x_0$.
 Note the appearance of the secular terms both $t$- and 
$x$-directions; these terms  were
 to be avoided in the RP method
  with the use of the solvability condition. 
 
Adding all the terms,  we finally get
\beq
\bfu(t;t_0)&=&\biggl[(\eps W(t_0,x_0)\bfU 
  + \eps ^3 \{A_1\Big(c_1(t-t_0+\delta)-
\frac{c_2}{2} D^{-1}(x^2 -x_0^2+\delta')\Big)\bfU \nonumber \\
 \ \ \ &\ & \ \ \ 
+ \sum _{\alpha\not=1}\frac{A_{\alpha}}{i\omega_0 - \lambda_0^{\alpha}}
\bfU_{\alpha}\}\biggl]{\rm e}^{i\omega_0t}
 + {\rm c.c.} + {\rm h.h.},
 \eeq
up to $O(\eps ^4)$.
 The RG/E equation\footnote{See section 3.} 
\beq
\frac{d {\bfu}}{d t_0}\Biggl\vert_{t_0=t}={\bf 0}, \ \ \ 
\frac{d {\bfu}}{d x_0}\Biggl\vert_{x_0=x}={\bf 0}, \ \ \ 
\eeq
 yields
\beq
\frac{\d W}{\d t}=\eps ^2c_1A_1 + O(\eps^3), 
\ \ \ \ D\frac{\d W}{\d x}=-\eps ^2xc_2A_1 +O(\eps^3).
\eeq
We remark that the seemingly vector equation is reduced to a scalar
 one.
Differentiating the second equation once again, we have
\beq
D\frac{\d ^2W}{\d x^2}=-\eps ^2c_2A_1 +O(\eps^3).
\eeq
 Here we have utilized the fact that $\d W/\d x= O(\eps^2)$.
 Noting that $c_1+c_2=1$, we finally reach
\beq
\frac{\d W}{\d t}- D\frac{\d ^2W}{\d x^2}&=&\eps ^2A_1, \nonumber \\  
\ \ &=& \eps^2\bigl[
  \chi \bfU ^{\ast}L_1\bfU W+ \{ 2\bfU^{\ast}M_0\bar{\bfU}{\bf V}_{+}
 +3\bfU^{\ast}N_0\bfU\bfU\bar{\bfU}\}\vert W\vert^2W\bigl] , 
\eeq
up to $O(\eps^3)$. 
This  is so called the time-dependent Ginzburg-Landau (TDGL) 
equation and
 coincides with the amplitude equation derived in the RP method\cite{kuramoto}.

We have seen that the RG method can reduce the dynamics of 
 a class of non-linear equations as  the RP method 
 can. Therefore it is needless to say that our method can be applied to
 the Brusselators\cite{brussel}, for instance, and leads to the same amplitude 
equations as the RP method\cite{kuramoto} does\cite{kunihiro4}.

\section{A brief summary and concluding remarks}

In this paper, we have shown that the RG method of Goldenfeld, Oono and 
 their collaborators can be 
 equally applied to vector equations, i.e.,
systems of ODE's and PDE's, as to scalar 
equations.\cite{goldenfeld1,goldenfeld2,kunihiro,kunihiro2}
We have  formulated the method
 on the basis of the classical thoery of envelopes, thereby
 completed the argument given in \cite{kunihiro,kunihiro2}. 
We have worked out for some examples of systems of ODE's, i.e.,
 the forced Duffing, the Lotka-Volterra and the Lorenz equation. 
 It has been also shown in a generic way that the method applied to 
 equations with a  bifurcation leads to the amplitude equations, such as
 the Landau-Stuart and 
 the (time-dependent) Ginzburg-Landau equation.

Then how about the phase equations\cite{kuramoto}?:
The phase equations describe another reduced dynamics.
The basis of the reduction of the dynamics by the phase equations lies in the
 fact that when a symmetry is broken, there appears 
a slow motion which is a classical counter part  of the Nambu-Goldstone boson
 in quantum field theory.
 We believe that if the phase equations are related to slow motions of the
 system at all, the RG method should also leads to the phase equations. 
  It is an interesting task to be done to show that it is the case.

There is another class of dynamics than those described by differential 
equations, i.e.,  difference equations or discrete maps.
 It is interesting  that a 
natural extension of the RG/E equation to  difference
 equations leads to a reduction of the dynamics.\cite{kunihiro3}
 This fact suggests that the RG method pioneered by Goldenfeld , Oono and
 their collaborators provides one of the  most promising candidate for 
  a general theory of
 the reduction of  dynamics, although it is certain that such a mechanical
 and general  method is often tedious in the actual calculations.\footnote{
It should be mentioned that there are other methods
 \cite{other1,other2} for the dynamical reduction
 as promising as the RG and RP method are.} 
As an application of the reduction of difference equations,
it will be interesting to see whether the coupled map lattice  
equations as systems of non-linear difference equations\cite{cml}
 can be  reduced to simpler equations by the RG method.  We hope that we can
 report about it in the near future.

\vspace{2.5cm}
\cl{\large{\bf Acknowledgements}}

This work is partly a reply to some people who asked 
 if the RG method could be applied to vector equations.
 The author acknowledges 
M. Davis and Y. Kuramoto for questions and comments for the previous
 papers\cite{kunihiro,kunihiro2}.
 He  also thanks  J. Matsukidaira and T. Ikeda for indicating the 
 significance of examining  vector equations.
 J Matsukidaira is gratefully acknowledged for useful comments  
in the earlist stage of this work.  He thanks M. Yamaguti and 
Y. Yamagishi for discussions on difference equations.
 He is indebted  to  R. Hirota,  H. Matano, 
  Y. Nishiura, J. Satsuma and  M. Yamaguti 
for their interest in this work.  He expresses his sincere thanks to M.
 Yamaguti for his encouragement.

{\large {\bf  Note added}}
 After submitting the paper, the author was informed that S. Sasa applied
 the RG method to derive phase equations in a formal way.
 The author is grateful to S. Sasa
 for sending me the TEX file({\tt patt-sol/9608008}) 
 of the paper before its publication.

\newpage
\setcounter{equation}{0}
\cl{\bf {\large Appendix A}}
\renewcommand{\theequation}{A.\arabic{equation}}

In this Appendix, we give a quick  review of Goldenfeld, Oono and 
their collaborators' prescription for the RG method. Then we summarize 
the problems to which a mthematical reasoning is needed in the author's 
 point of view.

We take the following simplest example to 
 show  their  prescription:
\beq
\frac{d^2 x}{dt^2}\ +\ \eps \frac{dx}{dt}\ +\ x\ =\ 0,
\eeq
where $\eps$ is supposed to be small. The exact solution reads
\beq
 x(t)= A \exp (-\frac{\eps}{2} t)\sin( \sqrt{1-\frac{\eps^2}{4}} t + \theta),
\eeq
where $A$ and $\theta$ are constant to be determined by an initial 
 condition.

  Now, let us blindly apply the perturbation theory  expanding $x$ as
\beq
x(t) = x_0(t) \ +\ \eps  x_1(t)\ +\ \eps ^2 x_2(t)\ +\ ... .
\eeq
The result is found to be\cite{kunihiro}
\beq
x(t; t_0)&=& A_0\sin (t +\theta_0) -\eps\frac{A_0}{2} (t -t_0)\sin(t+\theta_0)
  \nonumber \\ 
 \ \ \ & \ \ \ & +\eps^2\frac{A_0}{8}
\{ (t-t_0)^2\sin(t +\theta_0) - (t-t_0)\cos(t+\theta_0)\}
  + O(\eps^3).
\eeq

Now here come the crucial steps of the Goldenfeld et al's prescription:
\begin{description}
\item{(i)}
 First they introduce a dummy time $\tau$ which is close to $t$, and 
 ``renormalize" $x(t; t_0)$ by writing
 $t - t_0 = t-\tau +\tau - t_0$;
\beq
x(t, \tau)&=& A(\tau)\sin (t +\theta(\tau)) -\eps\frac{A(\tau)}{2}
 (t -\tau)\sin(t+\theta(\tau))
  \nonumber \\ 
 \ \ \ & \ \ \ & + \eps^2\frac{A(\tau)}{8}
\{ (t-\tau)^2\sin(t +\theta(\tau)) - (t-\tau)\cos(t+\theta(\tau))\}
  + O(\eps^3),
\eeq
with 
\beq
x(\tau, \tau)= A(\tau)\sin (\tau +\theta(\tau)).
\eeq
Here $A_0$ and $\theta_0$ have been multiplicatively renormalized to 
$A(\tau)$ and $\theta(\tau)$.

\item{(ii)} 
They observe that $\tau $ is an arbitrary constant introduced by hand, 
thus they claim that  
the solution $x(t, \tau)$ should not
depend on $\tau$; namely, $x(t, \tau)$ should satisfy the  equation
\beq
 \frac{d x(t, \tau)}{d \tau}=0.
\eeq
This  is similar to the RG equation in the field theory 
where $\tau$ corresponds to the 
 renormalization point $\tau$; hence the name of the RG method.

\item{(iii)}
 Finally they  impose another important but a mysterious condition that
\beq
\tau=t.
\eeq
\end{description}

From (ii) and (iii), one has
\beq
\frac{dA}{d\tau} + \frac{\eps}{2} A=0,  \ \ \ 
\frac{d\theta}{d\tau}+\frac{\eps^2}{8}=0,
\eeq
 which gives 
\beq
A(\tau)= \bar{A}{\rm e}^{-\eps\tau/2}, \ \ \ 
\theta (\tau)= -\frac{\eps^2}{8}\tau + \bar{\theta},
\eeq
where $\bar{A}$ and $\bar{\theta}$ are constant numbers.
 Thus, rewriting $\tau$ to $t$ in $x(\tau)$, one gets
\beq
x(t,t)= \bar{A}\exp(-\frac{\eps}{2} t)\sin((1-\frac{\eps ^2}{8})t + 
\bar{\theta}).
\eeq
They  identify $x(t,t)$ with the desired solution $x(t)$. Then one finds
 that the resultant $x(t)$ is an approximate but uniformly valid
 solution to Eq.(A.1).
In short, the solution obtained in the perturbation theory with 
 the local nature has been ``improved'' by the RG equation Eq.(A.7)
 to become a global solution.

But what have we done mathematically?
what is a mathematical meaning of the "renormalization''
 replacing $t_0$ with the extra dummy time $\tau$?
 Can't we avoid the "renormalization''
 procedure to solve a purely mathematical problem?
Why can we identify $x(t,t)$ with  the desired solution?; 
with $\tau $ being a constant, $x(t, \tau)$ can be a(n) (approximate)
 solution to Eq. (A.1), can't it?  In other words, when the operator $d/dt$
 hits the second argument of $x(t, t)$, what happens?  

In Ref.\cite{kunihiro}, it was shown  that  
the ``renormalization" procedure to introduce the extra dummy
 time $\tau$ is not necessary. 
Furthermore, it was clarified that   
the conditions (ii) and (iii) are   the ones to construct 
 the {\em envelope} of the family  of the local solutions 
obtained in the  perturbation theory;
 $x(t; t)$ is the envelope function of the
 family of curves given by $x(t; t_0)$ where $t_0$ parametrizes the 
 curves in the family. 
 Furthermore, it was shown that the envelope function $x(t,t)$  satisfies
 the orginal equations approximately but uniformly; the hitting of $d/dt$ 
on the second argument of $x(t, t)$ does not harm anything.
In short, the prescription given by Goldenfeld, Oono and their collaborators
 is not incorrect, but the reasoning for the prescription is given in 
\cite{kunihiro,kunihiro2} and will be more refined in the present
 paper.
In Ref.\cite{kunihiro2}, a simplification of the prescription and its 
 mathematical foundation is given for PDE's.

\newpage 
\cl{\bf {\large Appendix B}}
\setcounter{equation}{0}
\renewcommand{\theequation}{B.\arabic{equation}}

In this Appendix, we solve the forced Duffing equation without converting it
 to  a system.  It is easier to solve it in this  way than 
in the way shown in the text.

 We start with Eq. (2.6)
\beq
\ddot {z}+ 2\eps \gamma \dot{z}+ (1+\eps \sigma)z +
\frac{\eps h}{2}(3\vert z\vert^2z +{z^{\ast}}^3)= \eps f\e^{it},
\eeq
where $\eps$ is small.

Expanding $z$ as
\beq
z=z_0+\eps z_1 +\eps ^2z_2 + \cdots,
\eeq
one gets for $z$ in the perturbation theory 
\beq
z(t; t_0)= W(t_0)\e ^{it}+
\eps (t-t_0)\{f- W(\sigma + 2i\gamma) - \frac{3h}{2}\vert W\vert^2W\}
\e ^{it} + \eps \frac{1}{16}{W^{\ast}(t)}^3\e^{3it} + O(\eps^2).
\eeq
Note that there exists a secular term in the first order term.

The RG/E equation reads\cite{kunihiro}
\beq
\frac{d z}{d t_0}=0
\eeq
with $t_0=t$, which leads to
\beq
\dot{W}=
 -\eps(\sigma +2i\gamma)W - \frac{3}{2}\eps h\vert W\vert ^2W+ \eps f 
\eeq
 up to $O(\eps ^2)$.  Here we have discarded terms such as $\eps dW/dt$,
 which is $O(\eps ^2)$ because $dW/dt=O(\eps)$.
The resultant equation for the amplitude is the Landau-Stuart equation for the 
Duffing equation. The envelope is given
\beq
z_E(t)=z(t; t_0=t)=
 W(t)\e ^{it} + \frac{\eps}{16} {W^{\ast}}^3\e^{3it} + O(\eps^2).
\eeq
We identify $z_E(t)$ with a global solution of Eq.(B.2), and 
$x(t)={\rm Re}[z_E]$ and $y(t)={\rm Im}[z_E]$ are solutions to 
 Eq.(B.1).  As shown in the text, $\forall t$, 
$z_E(t)$ satisfies Eq.(B.2) uniformly up to $O(\eps ^2)$.

\newpage 
\newcommand{\NG}{N. \ Goldenfeld}
\newcommand{\YO}{Y.\ Oono}

\end{document}